%% file: Paper-0531.tex
\pdfoutput=1
\RequirePackage{amsmath}
\documentclass[runningheads]{llncs}
\usepackage{lipsum}
\usepackage{ulem}
\usepackage{bbm}

\usepackage{bm}
\usepackage{mathtools}
\usepackage{amssymb}
\usepackage{multicol,multirow}
\usepackage[noend]{algpseudocode}
\usepackage{algorithmicx,algorithm}
\usepackage{parcolumns}
\usepackage{booktabs}
\usepackage{xcolor}

\usepackage[T1]{fontenc}
\usepackage{graphicx}
\newcommand{\myparagraph}[1]{\subsubsection*{\textbf{#1}}}

\begin{document}
\title{Slice-Consistent 3D Volumetric\\Brain CT-to-MRI Translation with\\2D Brownian Bridge Diffusion Model}
\titlerunning{Slice-Consistent 3D CT2MRI with 2D Brownian Bridge Diffusion Model}

\author{
Kyobin Choo\inst{1} \and
Youngjun Jun\inst{2} \and
Mijin Yun\inst{3} \and
Seong Jae Hwang\inst{2}
}

\authorrunning{K. Choo et al.}

\institute{
Department of Computer Science, Yonsei University, Seoul, Republic of Korea \and
Department of Artificial Intelligence, Yonsei University, Seoul, Republic of Korea \and
Department of Nuclear Medicine, Yonsei University College of Medicine, Seoul, Republic of Korea\\
\email{\{chu, youngjun, seongjae\}@yonsei.ac.kr; yunmijin@yuhs.ac
}}

\maketitle              

\input{tex/00_abs}

\input{tex/01_intro}

\input{tex/02_methods}

\input{tex/03_exp}

\input{tex/04_conc}

\input{tex/credit}

%
\bibliographystyle{splncs04}
\bibliography{Paper-0531}

\newpage
\input{tex/sup}

\end{document}

%% file: tex/00_abs.tex
\pdfoutput=1
\begin{abstract} 
In neuroimaging, generally, brain CT is more cost-effective and accessible imaging option compared to MRI. Nevertheless, CT exhibits inferior soft-tissue contrast and higher noise levels, yielding less precise structural clarity. In response, leveraging more readily available CT to construct its counterpart MRI, namely, medical image-to-image translation (I2I), serves as a promising solution. Particularly, while diffusion models (DMs) have recently risen as a powerhouse, they also come with a few practical caveats for medical I2I. First, DMs' inherent stochasticity from random noise sampling cannot guarantee consistent MRI generation that faithfully reflects its CT. Second, for 3D volumetric images which are prevalent in medical imaging, na\"ively using 2D DMs leads to slice inconsistency, e.g., abnormal structural and brightness changes. While 3D DMs do exist, significant training costs and data dependency bring hesitation. As a solution, we propose novel style key conditioning (SKC) and inter-slice trajectory alignment (ISTA) sampling for the 2D Brownian bridge diffusion model. Specifically, SKC ensures a consistent imaging style (e.g., contrast) across slices, and ISTA interconnects the independent sampling of each slice, deterministically achieving style and shape consistent 3D CT-to-MRI translation. To the best of our knowledge, this study is the first to achieve high-quality 3D medical I2I based only on a 2D DM with no extra architectural models. Our experimental results show superior 3D medical I2I than existing 2D and 3D baselines, using in-house CT-MRI dataset and BraTS2023 FLAIR-T1 MRI dataset.

\keywords{Image-to-image translation \and Volumetric image synthesis \and Brownian bridge diffusion model \and Brain CT-to-MRI}
\end{abstract}

%% file: tex/01_intro.tex
\pdfoutput=1
\section{Introduction}
\label{sec:intro}

Computed Tomography (CT) and Magnetic Resonance Imaging (MRI) scans are crucial in diagnosing and treating various medical conditions \cite{MRI1,MRI_CT1}, each with its own pros and cons. CT quickly delivers tissue conditions and attenuation maps but struggles with detailed soft-tissue imaging \cite{CT1}. Conversely, MRI offers outstanding soft-tissue clarity but comes with higher costs and longer acquisition times compared to CT \cite{mri_vs_ct}. This juxtaposition between CT and MRI sets the stage for exploring synthesized MRI from CT images in scenarios where only CT is available \cite{CT2MRI_1}. Obtaining MRI without additional scan enhances cost-efficiency and diagnostic accuracy, especially in practices like positron emission tomography/computed tomography (PET/CT) scan readings, where MRI's clear anatomical details improves brain scan interpretations \cite{PET_CT_MR}. However, CT scans present some anatomical details subtly due to their lower soft-tissue contrast and higher noise levels \cite{CT_has_noise}, presenting a challenge in synthesizing faithful MRI from CT. This task can be defined as a image-to-image translation (I2I), which has key implications as it enables critical medical imaging tasks to be addressed, such as super-resolution \cite{efficient_SR}, denoising \cite{diff_denoising}, and multi-modal synthesis \cite{3d_ldm_multi_modal_mri}.

\begin{figure}[t!]
    \centering
    \includegraphics[width=1.0\textwidth]{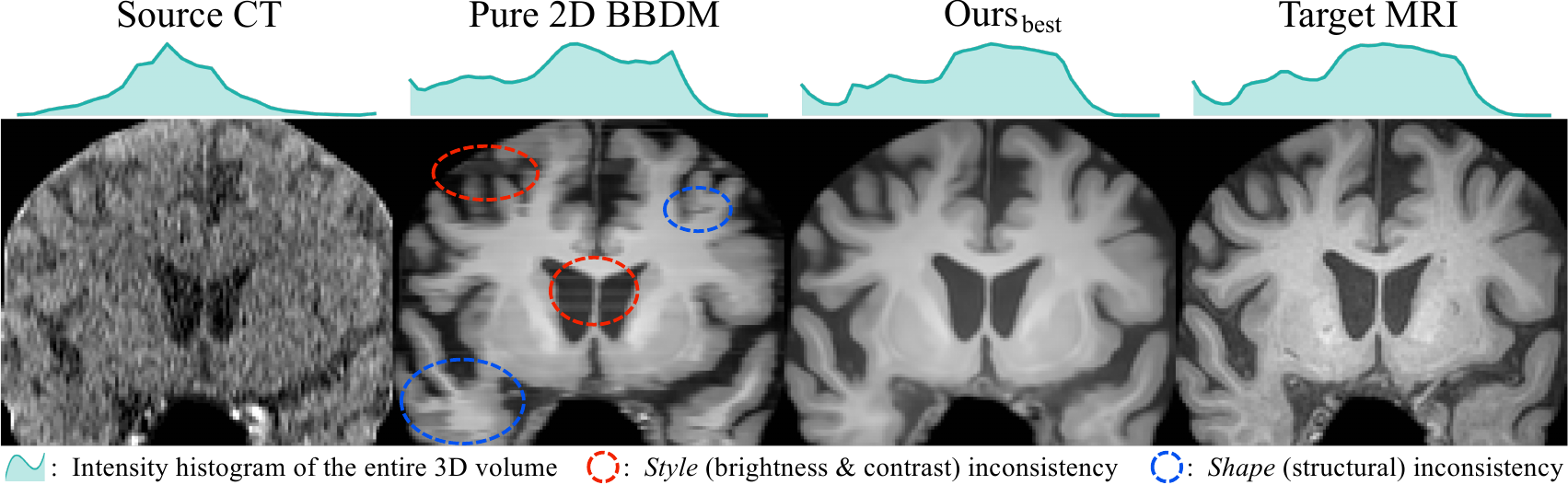}
    \caption{
    \textbf{Examples of slice inconsistency and resolved outcomes.}
    This figure displays the coronal view of volumes synthesized by two 2D BBDMs trained on axial slices. The pure multi-slice 2D BBDM exhibits severe slice inconsistency, with noticeable discontinuities in both style and shape across slices. Our method produces slice-consistent volumes and can adjust the intensity histogram (i.e., style).
    }
    \label{fig:motivation}
\end{figure}

Recently, diffusion models (DMs), a type of deep generative model, have been increasingly favored in various medical imaging tasks \cite{3d_ldm_multi_modal_mri,sure_diff,diff_denoising}, thanks to their ability to produce high-quality and diverse samples \cite{beat_gan,diffusionGAN}.
However, in medical I2I like CT-to-MRI translation, where it is essential for synthetic images to reliably reflect the structure of source images, DMs struggle to produce consistent outcomes due to their inherent stochasticity. Even using deterministic samplers like DDIM \cite{ddim}, their reliance on sampling from random noise introduces uncertainty, undermining reproducibility and medical reliability. Therefore, we performed medical I2I based on the Brownian Bridge Diffusion Model (BBDM \cite{bbdm}), which specifies the theoretical mapping between source and target data distributions through a diffusion process directly on the source image itself.

Meanwhile, medical imaging often involves 3D volumetric images. However, applying na\"{i}ve 3D models to high-resolution volumes is impractical, due to the massive computational, memory, and data requirements \cite{make_a_volume}. As a solution, approaches based on latent diffusion models (LDMs), which compress the latent space using an autoencoder to train 3D LDMs, have been proposed \cite{make_a_volume,3d_ldm_multi_modal_mri}. However, the image quality of LDM relies on its 3D autoencoder, which faces challenges in achieving satisfactory performance on high-dimensional data and is still costly for ordinary GPUs in memory usage \cite{3d_autoencoder_1,3d_autoencoder_2}.

An alternative of using 2D models for slice-wise generation exists. However, this leads to \textit{slice inconsistency} within a volume, as seen in Fig.~\ref{fig:motivation}, which can be defined in two aspects: \textit{style} (e.g., brightness and contrast) and \textit{shape} (e.g., structural discontinuities in 3D views).
In this paper, we aim to address this slice inconsistency issue in 2D models through two novel methods. \textbf{(1) Style Key Conditioning (SKC)}: We condition the DM with the MRI's histogram to enable it to learn the mapping between the histogram and the actual imaging style. By controlling the style of the generated slices through the histogram, we not only ensure style consistency but also enable the free manipulation of the target MRI's style. \textbf{(2) Inter-Slice Trajectory Alignment (ISTA)}: We ensure consistency in both style and shape by enabling adjacent slices to sample along a jointly agreed-upon trajectory, while still allowing for parallel slice-wise inference within each batch. Specifically, ISTA aggregates the multiple outputs from the model to \textit{co-predict} the direction of the next time step for each slice and deterministically \textit{corrects} the transition error from the co-prediction.

\paragraph{\textbf{Contributions.}}
Our main contributions are as follows: \textbf{(\romannumeral 1)} Through our proposed SKC and ISTA, we performed slice-consistent 3D volumetric brain CT-to-MRI translation using a 2D DM. \textbf{(\romannumeral 2)} By combining our approach with BBDM, we achieved fully deterministic and reliable medical I2I, with the target volume's structure and style dictated only by the source volume and histogram-based condition, respectively. \textbf{(\romannumeral 3)} Using in-house CT-MRI dataset and public FLAIR-T1 MRI dataset from BraTS2023, we demonstrate that proposed 2D-based method performs superior cross-modality I2I compared to existing 2D and 3D baseline models. To the best of our knowledge, this is the first study to achieve high-quality 3D medical I2I based only on a 2D DM with no extra architectural models like autoencoders. The code for our project can be accessed at \url{https://github.com/MICV-yonsei/CT2MRI}.

%% file: tex/02_methods.tex
\pdfoutput=1
\section{Methods}
In this section, we first briefly explore BBDM (Sec.~\ref{sec:method1}). Then, we detail the multi-slice training strategy with SKC for \textit{global} style consistency (Sec.~\ref{sec:method2}) and ISTA sampling technique for \textit{local} consistency both in style and shape (Sec.~\ref{sec:method3}).

\subsection{Preliminaries: 2D Brownian Bridge Diffusion Model}
\label{sec:method1}
BBDM redefines the diffusion model within the framework of a Brownian bridge to generalize the source distribution from Gaussian to any arbitrary distribution, making it suitable for I2I. Let $(\bm{x}, \bm{y})$ be any source and target image pair, then the forward Brownian bridge diffusion process at time step $t$ is defined based on $\bm{x}_0=\bm{x}$ (e.g., the target MRI slice) and $\bm{x}_T=\bm{y}$ (e.g., the source CT slice) as
\small
\begin{equation}
    q_{BB}(\bm{x}_t | \bm{x}_0, \bm{y}) = \mathcal{N}(\bm{x}_t; (1-m_t)\bm{x}_0 + m_t\bm{y}, \delta_t\bm{I}) \label{Eq-BB},
\end{equation}
\normalsize
where $\delta_t = 2s(m_t - m_t^2)$ is the variance with scale factor $s$ for the sampling diversity, and $m_t = t / T$. The Gaussian transition kernel of the forward Markov process $q_{BB}(\bm{x}_t | \bm{x}_{t-1}, \bm{y})$, with variance $\delta_{t|t-1} = \delta_t - \delta_{t-1}(1-m_t)^2/(1-m_{t-1})^2$, can be easily derived through Eq.~\eqref{Eq-BB}.
Similar to DDPM \cite{ddpm}, the forward process posterior $q_{BB}(\bm{x}_{t-1} | \bm{x}_t, \bm{x}_0, \bm{y})$ can be derived in closed form of Gaussian with variance $\Tilde{\delta}_t = \delta_{t|t-1}\cdot\delta_{t-1}/\delta_t$ and mean $\bm{\Tilde{\mu}}_t(\bm{x}_t, \bm{x}_0, \bm{y}) = c_{x t} \bm{x}_t+c_{y t} \bm{y}+c_{\epsilon t}(m_t(\bm{y}-\bm{x}_0)+\sqrt{\delta_t} \bm{\epsilon})$ for some coefficient $c_{xt}, c_{yt}$, and $c_{\epsilon t}$.
The reverse process is defined as $p_{\theta}(\bm{x}_{t-1} | \bm{x}_t, \bm{y}) = \mathcal{N}(\bm{x}_{t-1}; \bm{\mu}_{\theta}(\bm{x}_t, t), \Tilde{\delta}_t \bm{I})$, where $\bm{\mu}_{\theta}(\bm{x}_t, t)$ is the predicted mean by the model, and the variance $\Tilde{\delta}_t$ uses the same value as $q_{BB}(\bm{x}_{t-1} | \bm{x}_t, \bm{x}_0, \bm{y})$. The model parameters $\theta$ are trained to make the distribution of $q_{BB}(\bm{x}_{t-1} | \bm{x}_t, \bm{x}_0, \bm{y})$ and $p_{\theta}(\bm{x}_{t-1} | \bm{x}_t, \bm{y})$ identical, using a simplified evidence lower bound (ELBO) as
\small
\begin{align}
    \mathbb{E}_{\bm{x}_0, \bm{y}, \bm{\epsilon}}[c_{\epsilon t}||m_t(\bm{y}-\bm{x}_0) + \sqrt{\delta_t}\bm{\epsilon} - \bm{\epsilon}_{\theta}(\bm{x}_t, t)||^2_2], \label{BB-objective}
\end{align}
\normalsize
where $\bm{\epsilon} \sim \mathcal{N}(\boldsymbol{0}, \bm{I})$ is sampled random noise and $\bm{\epsilon}_{\theta}$ is the noise estimator model. A notable aspect in Eq.~\eqref{BB-objective} is that, the prediction $\bm{\epsilon}_{\theta}(\bm{x}_t,t)$ includes not just random noise $\bm{\epsilon}$. We have also observed that incorporating the guiding term $m_t(\bm{y}-\bm{x}_0) + \sqrt{\delta_t}\bm{\epsilon}$, which is the direction pointing to $\bm{x}_t$ from the predicted $\bm{x}_0$, improves performance over solely training the model to predict pure $\bm{\epsilon}$ practically.

\begin{figure}[t!]
    \centering
    \includegraphics[width=1.0\textwidth]{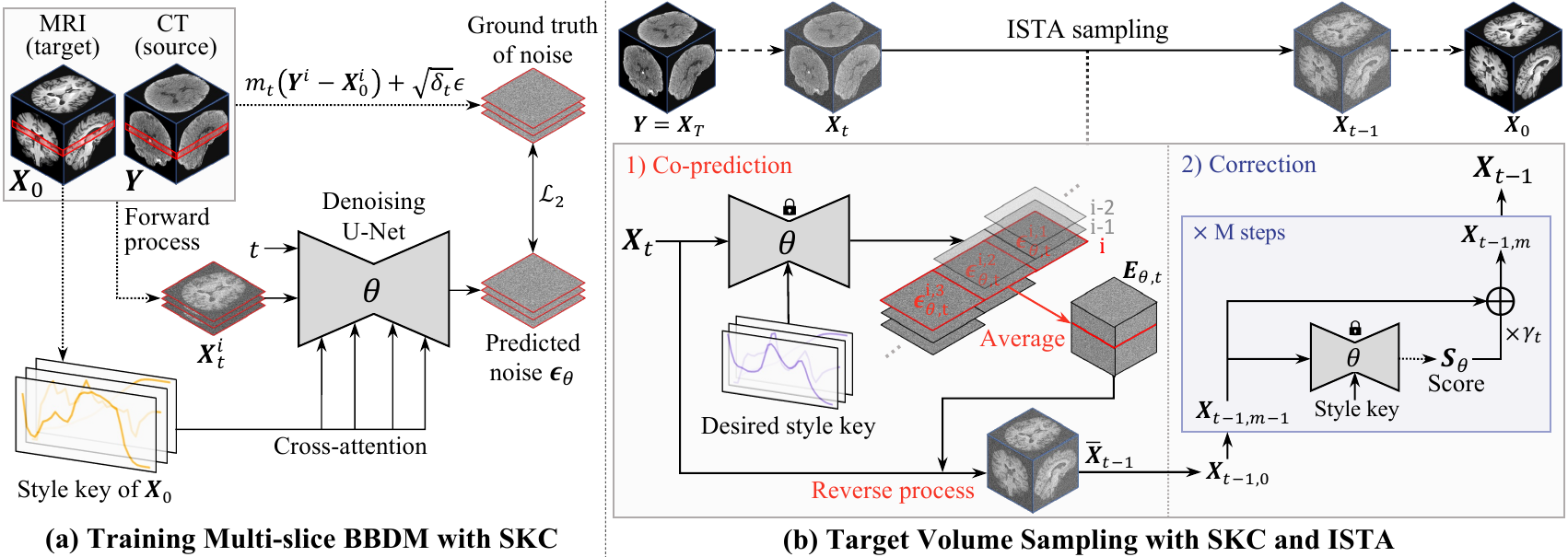}
    \caption{
    \textbf{Training and sampling scheme of the proposed methods.}
    \textbf{(a)} During the multi-slice BBDM training, a target histogram-based style key is injected into the U-Net.
    \textbf{(b)} Target volume sampling proceeds in the manner of the Predictor-Corrector method. During the \textit{co-prediction} phase, multiple $\bm{\epsilon}^{i,k}_{\theta,t}$ are employed to establish connections among the predicted slices within $\bm{\bar{X}}_{t-1}$. In the subsequent \textit{correction} phase, the co-predicted volume is refined through a score-guided deterministic process.
    }
    \label{fig:overview}
\end{figure}

\subsection{Training Multi-slice 2D BBDM with Style Key Conditioning}
\label{sec:method2}
When the 2D BBDM is na\"{i}vely trained slice-wise, the uneven brightness or contrast is observed globally in the synthetic MRI (Fig.~\ref{fig:motivation}), which we call \textit{global} style inconsistency. This occurs due to the target MRI in the training data exhibiting various styles, even for similar sources. To address global style inconsistency, we conditioned the model with a \textit{style key} that uniquely matches each target style, which also allows us to freely control the style of the generated slice. We discovered that the intensity histogram of an MRI volume can serve as an ideal style key because it vividly reflects imaging style while being invariant to an individual's anatomy. Consequently, we trained a 2D BBDM, conditioning on MRI histogram-based style keys (Fig.~\ref{fig:overview}a), employing a multi-slice input-output approach (1) to utilize the information from adjacent slices and (2) to compile multiple predictions for a single slice at each ISTA sampling step.

From now on, we define the notations as follows: $\bm{X},\bm{Y},\bm{X}_t \in \mathbb{R}^{Z \times H \times W}$ represent complete MRI, CT, and latent volumes, $\bm{x}^i, \bm{y}^i, \bm{x}_t^i  \in \mathbb{R}^{H \times W}$ are the $i^{th}$ slices of these volumes, and $\bm{X}^i, \bm{Y}^i, \bm{X}_t^i \in \mathbb{R}^{(2N+1) \times H \times W}$ denote sub-volumes of the $i^{th}$ index and its adjacent $2N$ slices (e.g., $\bm{X}^i=[\bm{x}^{i-N}, \bm{x}^{i-N+1}, \ldots, \bm{x}^{i+N}]$). Our objective for training multi-slice 2D BBDM modifies Eq.~\eqref{BB-objective} as follows:
\small
\begin{align}
    \mathbb{E}_{\bm{X}_0^i, \bm{Y}^i, \bm{\epsilon}}[c_{\epsilon t}||m_t(\bm{Y}^i-\bm{X}_0^i) + \sqrt{\delta_t}\bm{\epsilon} - \bm{\epsilon}_{\theta}(\bm{X}_t^i, \bm{c}_{SKC}, t)||^2_2],
\end{align}
\normalsize
where $\bm{\epsilon}$, $\bm{\epsilon}_{\theta}(\cdot)$ always share the same dimensions as the volume or slice input and $\bm{c}_{SKC} \in \mathbb{R}^{3 \times B}$ is a set of three 1D histograms of target MRI volume $\bm{X}$ (i.e., histogram, cumulative histogram, and histogram differential), each with $B$ bins. For testing, the histogram of any MRI with desired style can be used as $\bm{c}_{SKC}$. We set $N=1$ and $B=128$ which suffice our task with minimal extra cost.

\begin{figure}[t!]
    \centering
    \includegraphics[width=1\textwidth]{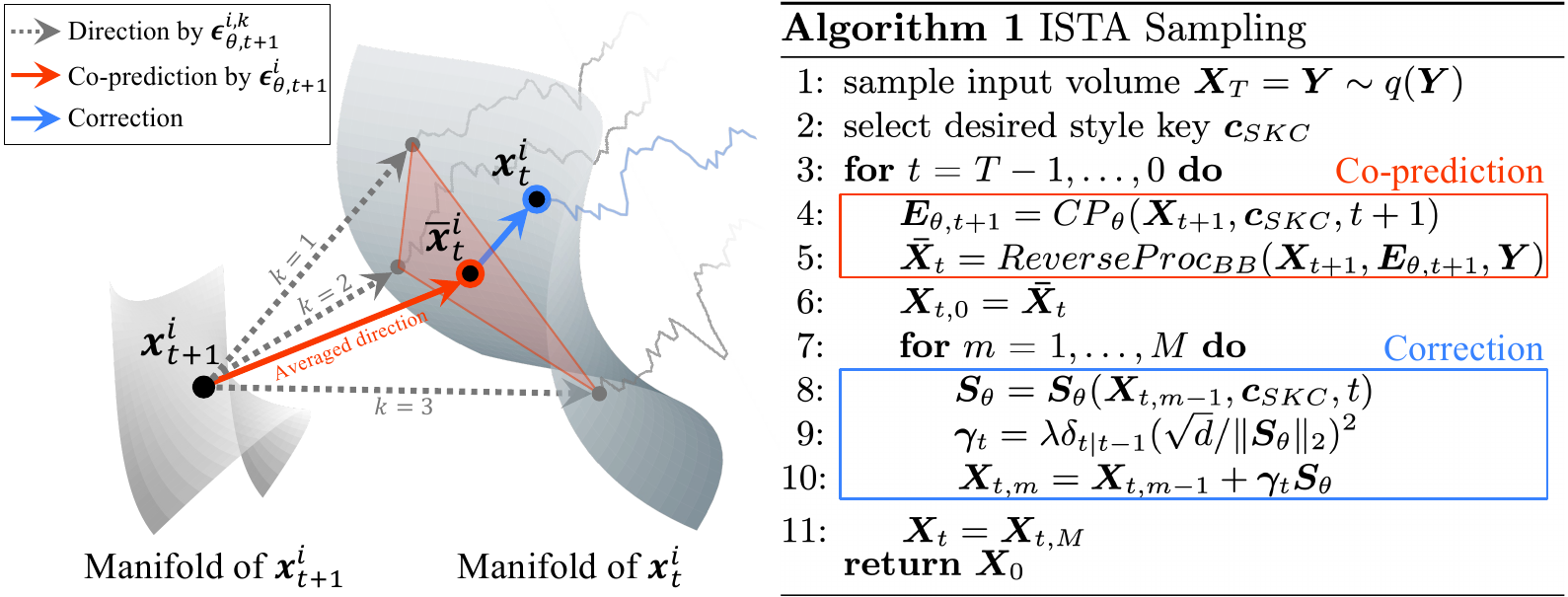}
    \caption{\textbf{Visualization of the latent space and algorithm for ISTA sampling.}
    The trained U-Net produces inconsistent outputs for multi-slice inputs that include the $i^{th}$ slice. The \textit{co-prediction} unifies the direction of these independent inferences, while the \textit{correction} aligns the co-predicted $\bar{\bm{x}}_{t}^i$ onto the manifold of $\bm{x}^i_{t}$.
    }
    \label{fig:ISTA}
\end{figure}

\subsection{Target Volume Sampling with Inter-slice Trajectory Alignment}
\label{sec:method3}
 While SKC addresses global style inconsistencies, it does not directly combat the \textit{local} slice inconsistency, which refers to style or shape discontinuities between adjacent slices. We resolve the local slice inconsistency by chaining all slices inside a volume exclusively within the 2D sampling process, without relying on any additional training beyond minimal multi-channel (i.e., 3-slice) burden. ISTA aligns the trajectories of adjacent slices during sampling to integrate independent inference outputs, ensuring it remains fully deterministic and parallel like BBDM. Similar to the Predictor-Corrector method in \cite{PC_method}, each ISTA sampling at time step $t$ involves two steps (Fig.~\ref{fig:overview}b and Fig.~\ref{fig:ISTA}): \textit{co-prediction} and \textit{correction}.

\paragraph{\textbf{Co-prediction.}}
ISTA sampling utilizes all predicted $\bm{\epsilon}_{\theta,t}^{i,k} \coloneqq \bm{\epsilon}_{\theta}(\bm{x}_t^{i,k}, \bm{c}_{SKC}, t) \in \mathbb{R}^{H \times W}$ for the $i^{th}$ slice of a volume obtained by the overlapping inference at time step $t$ $(k \in \{1, 2, \ldots, 2N+1\})$. First, we aggregate the $2N+1$ predictions for the $i^{th}$ slice into a single prediction $\bm{\epsilon}_{\theta,t}^i \coloneqq \mathbb{E}_k[\bm{\epsilon}_{\theta,t}^{i,k}]$. Subsequently, we concatenate all $\bm{\epsilon}_{\theta,t}^i$ to obtain the aggregated prediction volume $\bm{E}_{\theta,t}$. For an input volume $\bm{X}_t$, the process of co-predicting a single volume $\bm{E}_{\theta,t} \in \mathbb{R}^{Z \times H \times W}$ using multiple $i^{th}$ slice's outputs from the trained multi-slice model will be defined as $CP_{\theta}$:
\small
\begin{equation}
    \bm{E}_{\theta,t} = CP_{\theta}(\bm{X}_t, \bm{c}_{SKC}, t) \coloneqq  \left[\bm{\epsilon}_{\theta,t}^1, \bm{\epsilon}_{\theta,t}^2\, \ldots, \bm{\epsilon}_{\theta,t}^Z\right].
\end{equation}
\normalsize
Finally, according to the reverse process of BBDM, we obtain the volume $\bm{\bar{X}}_{t-1} \in \mathbb{R}^{Z \times H \times W}$ using the averaged prediction $\bm{E}_{\theta,t}$, $\bm{X}_t$, and source volume $\bm{Y}$.

\begin{figure}[t!]
    \centering
    \includegraphics[width=1.0\textwidth]{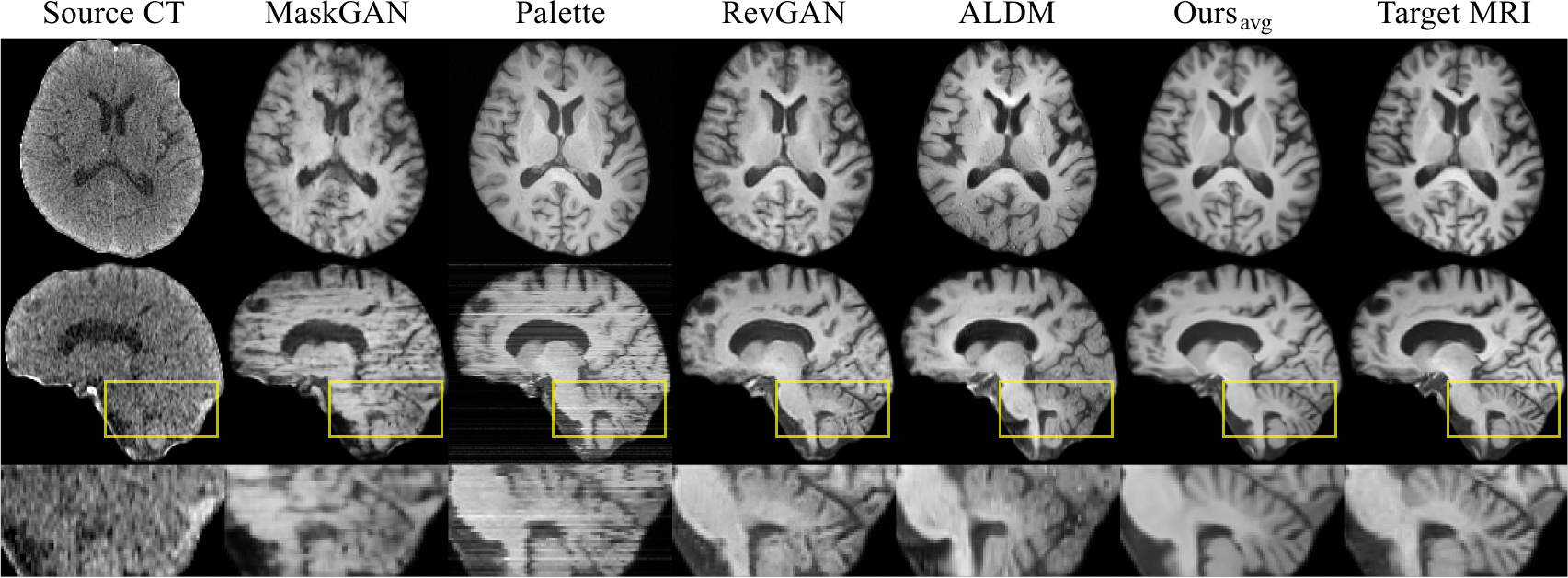}
    \caption{
    \textbf{Qualitative comparison with baselines (CT$\rightarrow$MRI)}
    }
    \label{fig:qual_result}
\end{figure}

\paragraph{\textbf{Correction.}}
However, $\bm{\bar{x}}_{t-1}^i$, having moved a time step via the simple average of  $\bm{\epsilon}_{\theta,t}^{i,k}$s (i.e., the averaged score at $\bm{x}_t^i$), does not guarantee its placement on the manifold of $\bm{x}_{t-1}^i$ (Fig.~\ref{fig:ISTA} left). To mitigate this transition error, inspired by the Langevin MCMC-based corrector in \cite{score_based_sde}, we propose a deterministic correction method with $M$ steps:
\small
\begin{equation}
   \bm{X}_{t,m} = \bm{X}_{t,m-1} + \lambda \delta_{t|t-1} (\sqrt{d} / \|\bm{S}_{\theta}\|_2)^2 \bm{S}_{\theta},
\end{equation}
\normalsize
where $\bm{X}_{t,0} = \bm{\bar{X}}_{t}$, $\bm{X}_t = \bm{X}_{t,M}$, the step size $\lambda$ is a hyperparameter, $d$ is the dimensionality of $\bm{x}_t^i$, and $\bm{S}_{\theta} = \bm{S}_{\theta}(\bm{X}_{t,m-1}, \bm{c}_{SKC}, t)$ is the score function \cite{score_based_sde} at $\bm{X}_{t,m-1}$. We removed the random noise $z$ of Langevin MCMC to ensure determinism, and applied an adaptive weighting for the magnitude of $\bm{S}_{\theta}$ and the variance. $\bm{S}_{\theta}$ can be derived similar to \cite{ddpm,sde_in_latent} on a volume basis as follows (Supp.~C):
\small
\begin{equation}
   \bm{S}_{\theta}(\bm{X}_t, \bm{c}_{SKC}, t) = -\frac{1}{\delta_t} \{m_t(\bm{X}_t - \bm{Y}) + (1-m_t){CP}_{\theta}(\bm{X}_t, \bm{c}_{SKC}, t)\}.
\end{equation}
\normalsize

%% file: tex/03_exp.tex
\pdfoutput=1
\section{Experiments}
\label{sec:experiments}

\begin{table*}[t]
  \centering
  \caption{\textbf{Quantitative comparison with baselines}}
    \begin{tabular*}{\textwidth}{@{\extracolsep{\fill}} l|cccccccc}
    \toprule[1pt]
    \multicolumn{2}{c}{} & \multicolumn{3}{c}{CT$\rightarrow$MRI (in-house)} &  \multicolumn{3}{c}{FLAIR$\rightarrow$T1 (BraTS)} \\
    \cmidrule(r){3-5} \cmidrule(r){6-8}
    \multicolumn{2}{c}{Methods} & NRMSE$\downarrow$ & PSNR$\uparrow$ & SSIM$\uparrow$ & NRMSE$\downarrow$ & PSNR$\uparrow$ & SSIM$\uparrow$ \\
    \midrule
    \multirow{2}{*}{ 3D } & RevGAN & {0.0577} & {25.344} & {0.8925} & {0.0824} & {22.073} & {0.8370}\\
                        & ALDM (200 step) & {0.0673} & {23.495} & {0.8474} & 0.0975 & 20.453 & 0.7921\\
    \midrule
    \multirow{3}{*}{ 2D } & MaskGAN & {0.0910} & {21.328} & {0.7421} & {0.1112} & {19.311} & {0.7222}\\
                        & Palette (1000 step) & {0.0811} & {21.877} & {0.4365} & {0.1495} & {16.974} & {0.4276} \\
                        & \textbf{Ours (100 step)} & \textbf{0.0515} & \textbf{26.666} & \textbf{0.9199} & \textbf{0.0808} & \textbf{22.579} & \textbf{0.8837} \\
    \bottomrule[1pt]
  \end{tabular*}
  \label{table:quant_result}
\end{table*}

\myparagraph{Datasets.}
We evaluate our method and baselines on in-house CT-MRI (T1) dataset and public FLAIR-T1 MRI dataset from BraTS2023 \cite{brats-baid2021rsna,brats-bakas2017segmentation,brats-bakas2017advancing,brats-menze2014multimodal}.
CT and MRI images from the in-house dataset were coregistered using SPM12 \cite{spm} and skull-stripped with SynthStrip \cite{synthstrip}. Both datasets were resampled to 1 mm isotropic voxels, cropped for background removal, and min-max normalized to a range of 0 to 1. Details of each dataset can be found in Supp.~A.

\myparagraph{Implementation Details.}
All 2D models used axial slices, and all DMs were trained with 1000 diffusion steps. Our method employed 100 DDIM sampling steps, and for ISTA sampling, we adjusted to 50 steps with $M=1$ correction for a fair comparison. For testing, the default style key of SKC is the averaged histogram of the entire training dataset. More details can be found in Supp.~A.

\myparagraph{Baselines.}
For comparison, we reproduced 2D and 3D GANs and DMs baselines for I2I, using default settings unless specified otherwise: (1) MaskGAN \cite{maskgan}, a cyclic GAN for unpaired 2D medical I2I, adapted for paired training; (2) Palette \cite{palette}, a conditional DM for 2D I2I; (3) RevGAN \cite{rev_gan}, a GAN with a reversible 3D backbone, also adapted for paired training; (4) ALDM \cite{3d_ldm_multi_modal_mri}, a 3D LDM framework, with model channels reduced to fit our GPU memory capacity.

\subsection{Evaluations}
\label{sec:exp2}

\myparagraph{Quantitative Results.}
We confirmed the effectiveness of our method against baselines on the in-house CT-MRI and BraTS FLAIR-T1 datasets (Table~\ref{table:quant_result}). For quantitative evaluation, normalized root mean square error (NRMSE), peak signal-to-noise ratio (PSNR), and structural similarity index measure (SSIM) were used. In examining baselines, we noted two points: DMs showed similar or lower metrics than GANs due to the diversity from their inherent stochasticity, and 3D models surpassed 2D models. Yet, our method, which is based on a 2D DM with small sampling step, outperformed all GANs and 3D models. This highlights the efficacy of (1) style uniformization through SKC, (2) shape and style consistency via ISTA, and (3) Brownian bridge-based stable mapping.

\begin{table*}[t!]
  \centering
  \caption{\textbf{Quantitative results of ablation studies.}} 
    \begin{tabular*}{\textwidth}{@{\extracolsep{\fill}} lcccccccc}
    \toprule[1pt]
     & \multicolumn{3}{c}{CT$\rightarrow$MRI (in-house)} &  \multicolumn{3}{c}{FLAIR$\rightarrow$T1 (BraTS)} \\
    \cmidrule(r){2-4} \cmidrule(r){5-7}
    Methods & NRMSE$\downarrow$ & PSNR$\uparrow$ & SSIM$\uparrow$ & NRMSE$\downarrow$ & PSNR$\uparrow$ & SSIM$\uparrow$ \\
    \midrule
    Pure BBDM  & 0.0575  & 25.359 & 0.8912 & 0.2232 & 16.639 & 0.7917\\
    +$\text{SKC}_{\text{colin}}$  & 0.0619 & 25.285 & 0.9004 & 0.0855 & 22.223 & 0.8806\\
    +$\text{SKC}_{\text{avg}}$  & 0.0517 & 26.443 & 0.9157 & 0.0820 & 22.409 & 0.8763\\
    +ISTA  & 0.0525  & 26.551 & 0.9167 & 0.1787 & 17.652 & 0.8054\\
    +$\text{SKC}_{\text{avg}}$, $\text{ISTA}_{\text{cp-only}}$ & 0.0515 & 26.664 & 0.9199 & 0.0809 & 22.578 & 0.8837 \\
    +$\text{SKC}_{\text{avg}}$, ISTA ($\text{Ours}_{\text{avg}}$)& \textbf{0.0515} & \textbf{26.666} & \textbf{0.9199} & \textbf{0.0808} & \textbf{22.579} & \textbf{0.8837} \\
    +$\text{SKC}_{\text{best}}$, ISTA ($\text{Ours}_{\text{best}}$)& \textbf{0.0454} & \textbf{27.589} & \textbf{0.9252} & \textbf{0.0625} & \textbf{24.644} & \textbf{0.8938} \\
    \bottomrule[1pt]
  \end{tabular*}
  \label{table:ablation}
\end{table*}

\myparagraph{Qualitative Results.}
Our method generated high-quality MRI volumes that are very similar to the target MRI, outperforming baselines (Fig.~\ref{fig:qual_result} \& Supp.~B). Compared to 3D models, 2D baseline models showed severe slice inconsistency in sagittal views. However, our 2D-based method excelled in slice consistency and surpassed 3D models in anatomical clarity across all views. Furthermore, our method can produce volumes that reflect various desired styles (Supp.~E).

\myparagraph{Ablation Studies.}
We conducted experiments on various options for SKC and ISTA (Table~\ref{table:ablation} \& Supp.~D).
Firstly, we set three options for SKC: the default averaged histogram ($\text{SKC}_{\text{avg}}$), the histogram of the public Colin 27 Average Brain \cite{colin} ($\text{SKC}_{\text{colin}}$), and the histogram similar to the target ground truth from the training data ($\text{SKC}_{\text{best}}$). The performance of $\text{SKC}_{\text{avg}}$ over pure BBDM and the elimination of style inconsistency indicate that SKC enables slice generation with uniform style. The performance of $\text{SKC}_{\text{colin}}$ may vary depending on the similarity between Colin's style and the test data, but it outperformed the pure BBDM and produced volumes with the desired styles (Supp.~E). The marked improvement with $\text{Ours}_{\text{best}}$ highlights our method's potential to accurately produce volumes close to the ground truth when provided with an exact style key for the target.
Secondly, two options were set for ISTA: with correction (ISTA, 50 step, $M=1$) and co-prediction only ($\text{ISTA}_{\text{cp-only}}$, 100 step, $M=0$). ISTA, $\text{ISTA}_{\text{cp-only}}$, and without ISTA showed superior performance in that order, proving the effect of both local slice consistency via co-prediction and error reduction via correction.
Ultimately, optimal performance was achieved with both SKC and ISTA.

%% file: tex/04_conc.tex
\pdfoutput=1
\section{Conclusion and Discussion}
\label{sec:conclusions}
We have successfully achieved slice-consistent and high-quality 3D volumetric brain CT-to-MRI translation with 2D BBDM, enhancing medical reliability through fully deterministic generation. Our proposed methods, SKC and ISTA, tackle the slice inconsistency issue from different perspectives and hold broad applicability in medical volume synthesis tasks. These fundamental approaches have significant potential for future expansion. SKC, which allows for control of the target imaging style based on the histogram, can be particularly utilized in addressing domain gap issues in medical imaging fields, such as multi-site MRI harmonization. Concurrently, ISTA can serve as a foundational method that enables iterative generative models, such as diffusion models, to establish connectivity in a higher-dimensional space than the input dimension.

%% file: tex/credit.tex
\pdfoutput=1
\begin{credits}
\subsubsection{\ackname} This work was supported in part by the IITP 2020-0-01361 (AI Graduate School Program at Yonsei University), NRF RS-2024-00345806, NRF RS-2023-00262002 funded by Korean Government (MSIT), and a grant of the Korea Dementia Research Project through the Korea Dementia Research Center (KDRC), funded by the Ministry of Health \& Welfare and Ministry of Science and ICT, Republic of Korea (RS-2022-KH127174).

\subsubsection{Disclosure of Interests.} The authors have no competing interests.

\end{credits}

%% file: tex/sup.tex
\pdfoutput=1
\begin{center}
\LARGE \textbf{Supplementary Materials}
\bigskip \\
\bigskip
\end{center}

\renewcommand\thesection{\Alph{section}.} 

\noindent\textbf{A. Experimental Details}
\begin{table*}[h]
    \centering
    \begin{minipage}[b]{0.48\textwidth} 
        \caption{\textbf{Dataset details}}
        \centering
        \scriptsize
        \renewcommand{\arraystretch}{1.168} 
        \scalebox{1.09}{
            \begin{tabular*}{0.87\linewidth}{@{\extracolsep{\fill}} ccc}
            \toprule[1pt]
             & CT-MRI & FLAIR-T1 \\
            \midrule
            {\#total}  & {219}  & {1470} \\
            {\#train}  & {197}  & {1251} \\
            {\#test}  & {22}  & {219} \\
            {shape}  & {160$\times$160$\times$160}  & {176$\times$176$\times$160} \\
            {voxel size}  & \multicolumn{2}{c}{1$\times$1$\times$1$mm^3$} \\
            \bottomrule[1pt]
            \end{tabular*}
        }
    \end{minipage}
    \hfill
    \begin{minipage}[b]{0.48\textwidth}
        \caption{\textbf{Implementation details}}
        \centering
        \scriptsize
        \scalebox{1.09}{
            \begin{tabular*}{0.87\linewidth}{@{\extracolsep{\fill}} ccc}
            \toprule[1pt]
             & CT$\rightarrow$MRI & FLAIR$\rightarrow$T1 \\
            \midrule
            {batch size}  & {16}  & {8} \\
            {\#iterations}  & {100,000}  & {150,000} \\
            {$\lambda$}  & {0.5}  & {0.05} \\
            {time step}  & \multicolumn{2}{c}{1000} \\
            {sampling step}  & \multicolumn{2}{c}{100} \\
            {GPU}  & \multicolumn{2}{c}{NVIDIA RTX A6000} \\
            \bottomrule[1pt]
            \end{tabular*}
        }
    \end{minipage}
\end{table*}

\noindent\textbf{B. Additional Qualitative Comparison}
\begin{figure}[h]
    \centering
    \includegraphics[width=1.0\textwidth]{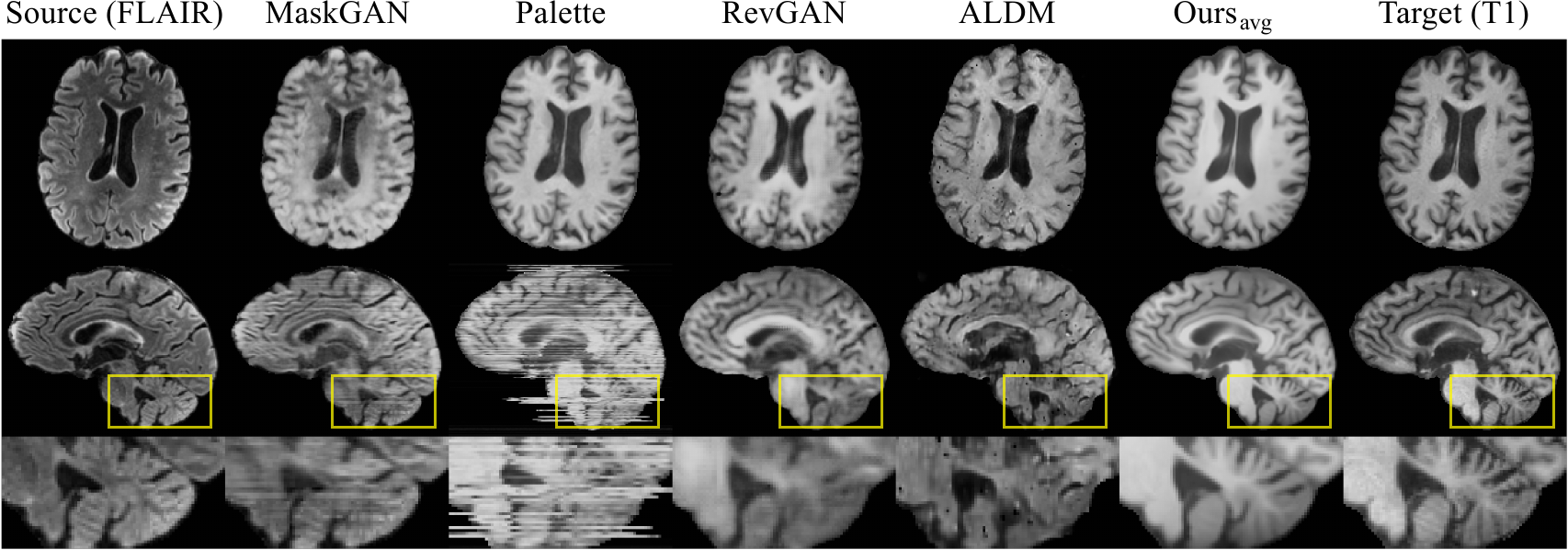}
    \caption{\textbf{Qualitative comparison with baselines (FLAIR$\rightarrow$T1)}
    }
\end{figure}

\noindent\textbf{C. Derivation of Score Function $\bm{S}_{\theta}(\bm{X}_t, \bm{c}_{SKC}, t)$ for $\theta$ Trained as $\theta=\theta^*$}
\bigskip
{\small
\begin{align*}
&\text{Since } \bm{E}_{\theta^*,t}=CP_{\theta^*}(\bm{X}_t, \bm{c}_{SKC}, t) = m_t(\bm{Y} - \bm{X}_0) + \sqrt{\delta_t}\bm{\epsilon}_t, \text{ where } \bm{\epsilon}_t \in \mathbb{R}^{Z\times H \times W},\\
& CP_{\theta^*}(\bm{X}_t, \bm{c}_{SKC}, t) = \bm{X}_t - \bm{X}_0 \; (\because \bm{X}_t = (1-m_t)\bm{X}_0 + m_t\bm{Y} + \sqrt{\delta_t}\bm{\epsilon}_t). \\
&\text{Then, } CP_{\theta^*}(\bm{X}_t, \bm{c}_{SKC}, t) =  m_t[\bm{Y} - \{\bm{X}_t - CP_{\theta^*}(\bm{X}_t, \bm{c}_{SKC}, t)\}] + \sqrt{\delta_t}\bm{\epsilon}_t \\
&\; \Longleftrightarrow \bm{\epsilon}_t  = \frac{1}{\sqrt{\delta_t}} \{m_t(\bm{X}_t - \bm{Y}) + (1-m_t)CP_{\theta^*}(\bm{X}_t, \bm{c}_{SKC}, t)\}. \\
&\text{Thus, } \bm{S}_{\theta^*}(\bm{X}_t, \bm{c}_{SKC}, t) = \nabla_{\bm{X}_t} \log q_{BB}(\bm{X}_t \mid (1-m_t)\bm{X}_0 + m_t\bm{Y}, \bm{c}_{SKC}) \\
&\; =-\frac{[\bm{X}_t - \{(1-m_t)\bm{X}_0 + m_t\bm{Y}\}]}{\delta_t} = -\frac{\bm{\epsilon}_t}{\sqrt{\delta_t}} \text{  (cf. } \nabla_{\tilde{\bm{x}}} \log q_\sigma(\tilde{\bm{x}} \mid \bm{x}) = -\frac{\tilde{\bm{x}}-\bm{x}}{\sigma^2} \text{ in \cite{NCSN}),} \\
&\; \text{since } \bm{X}_t \text{ is the Gaussian perturbation of } (1-m_t)\bm{X}_0 + m_t\bm{Y}. \\
&\therefore \bm{S}_{\theta^*}(\bm{X}_t, \bm{c}_{SKC}, t) = -\frac{1}{\delta_t} \{m_t(\bm{X}_t - \bm{Y}) + (1-m_t)CP_{\theta^*}(\bm{X}_t, \bm{c}_{SKC}, t)\}.
\end{align*}
\normalsize
}
\clearpage
\newpage

\noindent\textbf{D. Qualitative Results of Ablation Study}
\begin{figure}[h]
    \centering
    \includegraphics[width=1.0\textwidth]{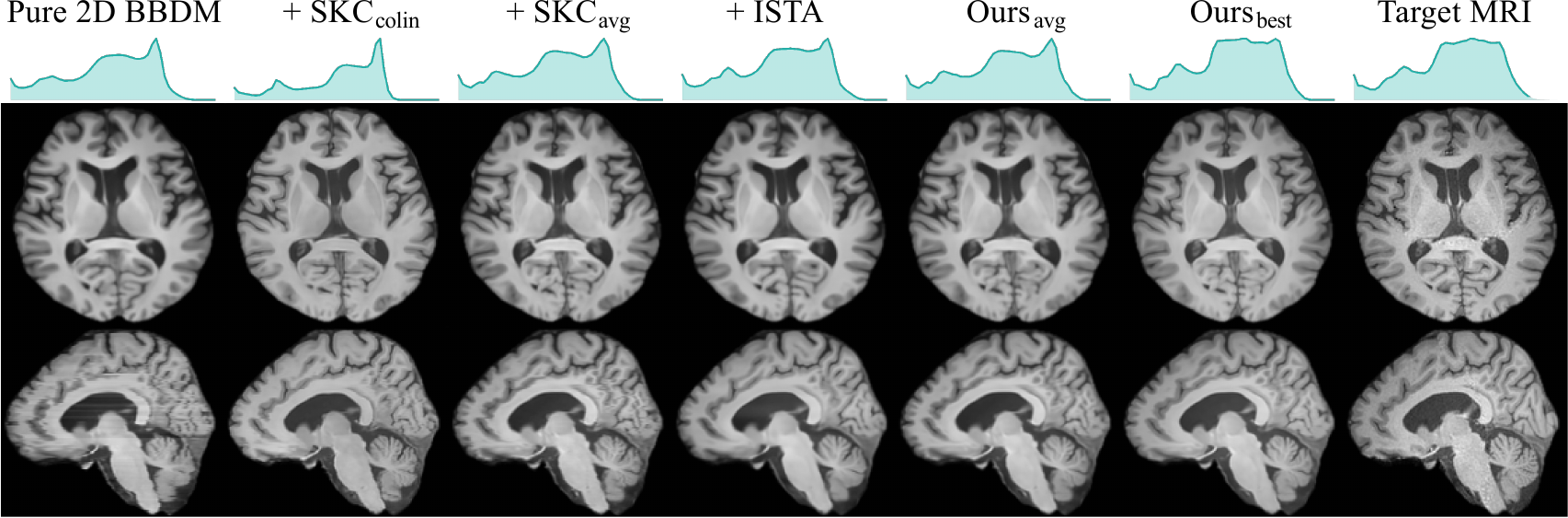}
    \caption{\textbf{Qualitative comparison with baselines (CT$\rightarrow$MRI)}
    }
\end{figure}

\noindent\textbf{E. Additional Study for Style Key Conditioning}
\begin{figure}[h]
    \centering
    \includegraphics[width=1.0\textwidth]{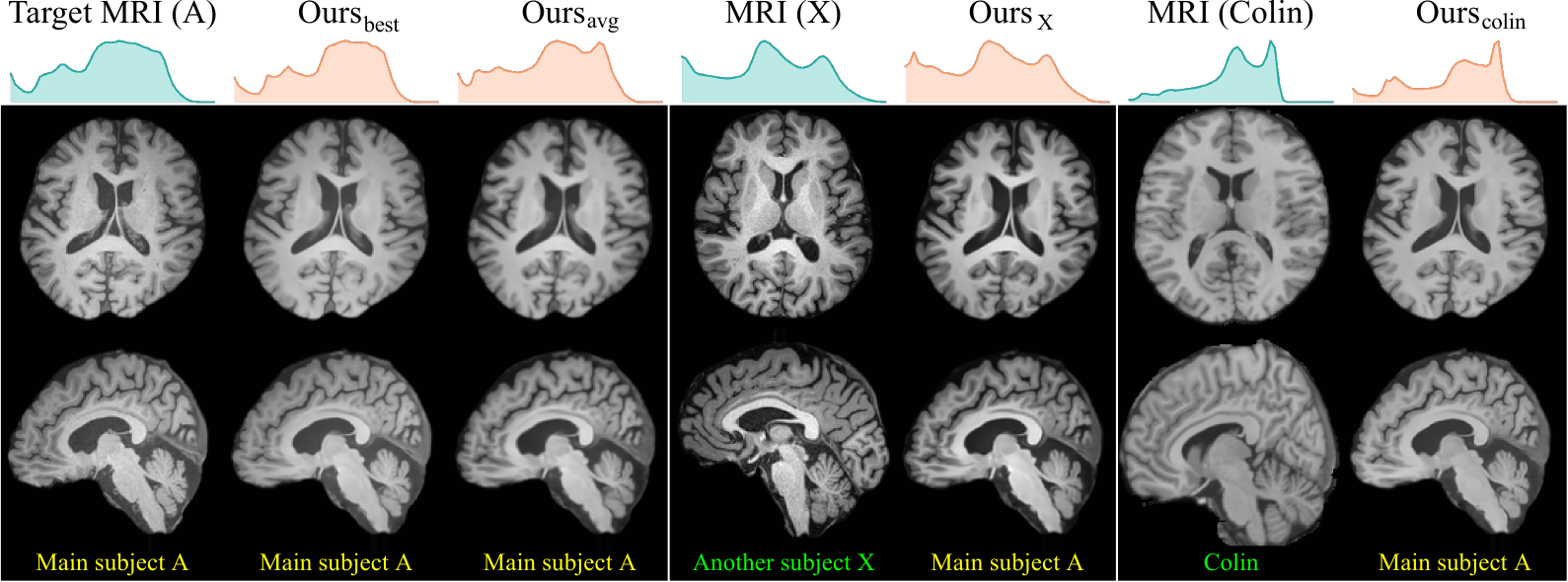}
    \caption{\textbf{Qualitative results for various style keys (CT$\rightarrow$MRI).}
    Figure shows real MRIs of main subject A, another subject X, and Colin 27 Average Brain (public access), along with synthetic MRIs of subject A. Both $\text{Ours}_{\text{X}}$ and $\text{Ours}_{\text{colin}}$ present synthetic MRIs of subject A, each transformed into the respective styles of X and Colin. From the analysis of the intensity histogram presented above, we observe that $\text{Ours}_{\text{X}}$ closely matches the MRI histogram of subject X, while $\text{Ours}_{\text{colin}}$ aligns with that of Colin. This demonstrates that the style of MRIs can be effectively controlled using SKC.
    }
\end{figure}